\documentclass[%
 reprint,
groupedaddress,
 amsmath,amssymb,
 aps,
prb,
]{revtex4-1}
\usepackage[cp1251]{inputenc}
\usepackage{graphicx}
\usepackage{dcolumn}
\usepackage{bm}

\begin{document}

\title{Magnetic Interaction in Doped $2D$ Perovskite Cuprates with Nanoscale Inhomogeneity: Lattice Nonlocal Effects vs Superexchange}
\author{Vladimir A. Gavrichkov}
\email[]{gav@iph.krasn.ru}
\affiliation{Kirensky Institute of Physics, Akademgorodok 50, bld.38, Krasnoyarsk, 660036 Russia}

\author{Semeon I. Polukeev}
\affiliation{Kirensky Institute of Physics, Akademgorodok 50, bld.38, Krasnoyarsk, 660036 Russia}

\date{\today}

\begin{abstract}

We have studied the superexchange interaction   $J_{ij}$ in doped $2D$ cuprates. The AFM interaction strongly depends on the state of the lattice of a CuO$_2$ layer surrounded by two LaO  rock salt layers. In a static $U$  and $D$  stripe nanostructure, the homogeneous AFM interaction is impossible due to the  $U/D/U...$  periodic stripe sequence and $T_N=0$.  In a dynamic stripe nanostructure, the ideal CuO$_2$  layer with nonlocal effects and the homogeneous AFM interaction are restored.  However the interaction $J_{ij}$  decreases by the exponential factor due to partial dynamic quenching. The meaning of the transition from the dynamic to the static cases lies into the spontaneous $\theta$-symmetry breaking with respect to the rotation of all the tilted  CuO$_6$ octahedra by an orientation angle $\delta \theta  = n \cdot \left( {45^\circ } \right)$ (where $n=1\div4$) in the $U$ and $D$ stripe nanostructure of the CuO$_2$ layer.
Moreover, the structural features help to study various experimental data on the charge inhomogeneity, Fermi level pinning in the p type cuprates only and a time reversal symmetry breaking from a unified point of view.

\end{abstract}


\maketitle

\section{\label{sec:intr}Introduction\\}

The unique functionality of several materials with perovskite structure, like cuprates,~\cite{High2017, Alex2017, Benedek1993, Bianconi2001} can be tuned by atomic substitutions, tolerance factor, misfit strain and pressure which control structural tilts and nanoscale phase separation.
The scanning tunneling microscopy and spectroscopy (STM and STS)~\cite{Zhao2019, Maggio2010, Fisher2007, Silva2014, Ma2009} and advanced experimental X-ray methods~\cite{Lanzara1997, Campi2015, Comin2016, Mitrano2019, Campi2021} in a wide range of temperature and doping unambiguously indicate that there is a clear connection between the multiscale stripe texture and the quantum coherence of quasiparticles in $2D$ perovskite high-Tc superconductors. The stripe charge nanostructure is also accompanied by spin inhomogeneity ~\cite{Zaanen1989,Machida1989, Tranquada1995, Fujita2004, Hucker2011, Fujita2012, Wen2019, Miao2021, Ma2021}, and the high-temperature superconductivity (HTSC)  is believed to originate from magnetic spin excitations that bind Cooper pairs ~\cite{Anderson1987, Scalapino1995, Dagotto1994, Izyumov1999}. The magnetic interaction in parent AFM $2D$ cuprates is the well-known superexchange interaction.~\cite{Anderson1959} The AFM interaction   between the nearest spins of  Cu$^{2+}$  ions in the CuO$_2$  layer is strong $\left( { \sim 0.146eV} \right)$~\cite{Coldea2001}, being  much stronger than the interplanar exchange which is mainly responsible for long-range magnetic ordering observed in undoped cuprates. For La$_2$CuO$_4$ (LCO), the  Neel temperature is $T_N  \approx 300K$. In the traditional picture of BCS superconductivity, magnetism destroys Cooper pairs. However, superconductivity in these HTSCs develops from a "bad metal", whose resistivity is higher than that of BCS superconductors, and the AFM magnetism itself is related to the initial "bad metal" state, where superconductivity occurs when long-range magnetic ordering is suppressed at a relatively weak hole doping $\left( {x \sim 0.01} \right)$. Indeed, in the resonating-valence bond (RVB) approach and $t-J$ model,~\cite{Anderson1987, Bascaran1987, Gros1987, Kotliar1988, Suzumura1988},  HTSC emerges due to the condensation of hole pairs, induced by the exchange interaction $J_{ij}$.  Pairing of holes in this mechanism is caused by the interband transfer of quasiparticles in the CuO$_2$ layer with short-range AFM ordering. Retardation effects for this mechanism are insignificant. Another view on the HTSC state in the cuprates (see for example ~\cite{Prelovchek2005, Plakida2003}) is that AFM spin fluctuations become particularly longer ranged and soft at low hole doping.  In fact, there is a resonant magnetic mode ~\cite{Rossat1991, Fong2000, Dai2001} within the HTSC phase, where the scenario of spin fluctuations represents the lowest bosonic mode relevant for the d-wave HTSC pairing in the energy range $\sim J_{ij}$.\\

If the magnetic interaction is a reliable candidate for being used as a glue for holes~\cite{Scalapino2012}, then how will it change in the inhomogeneous nanostructure of doped 2D cuprates? Actual research on the properties of a quantum spin liquid in the double perovskite oxides with the general formula of A$_2$Me$_2$O$_6$ ~\cite{Vasala2012, Katukuri2020, Thakur2021} does not provide a direct answer to this question, since there is no anion substitution in the CuO$_2$ layer of HTSC cuprates. However, the dynamical charge and spin nanoinhomogeneous is identified and discussed in current neutron and STM experiments ~\cite{Li2019, Du2020, Tsvelik2019, Chen2022}. Based on the large isotope effect in high-Tc superconductors with the stripe nanostructure~\cite{Lanzara1999, Rubio2000, Suryadijaya2005}, we will study superexhange interaction taking into account the phonon nature of the stripe structure in high-Tc cuprates. It will be shown that the charge and spin nanoscale phase separation can be derived at least qualitatively in the lattice approach, and the homogeneous interaction $J_{ij}$  is restored at a certain hole concentration corresponding to an equal-dimensional periodic stripe structure of a linear or chess type. Furthermore, this is a good reason to test your feelings in accordance with the opinion: "...there remains the nagging feeling that even if the basic pairing mechanism ultimately arises from the short-range antiferromagnetic correlations in the   layer, some important ingredient may be missing from the theory we have described." (see the review ~\cite{Scalapino1995}). In conclusion, we discuss the observable consequences from the lattice concept of the stripe nature.

\section{\label{sec:II} Color approach to nanoscale inhomogeneity and $D$ and $U$ stripe structure\\}
\begin{figure}
\includegraphics{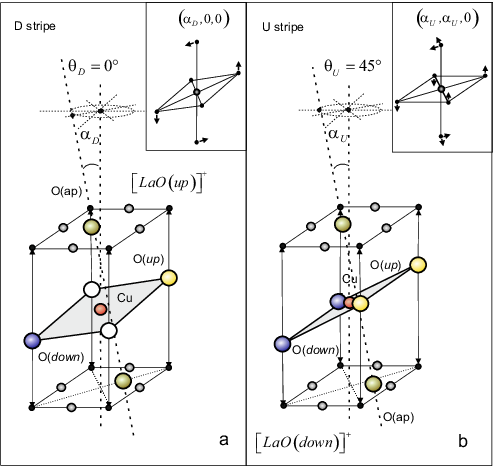}
\caption{\label{fig:1} View of the tilted  CuO$_6$ octahedra of the $D$ (a) and $U$ (b) stripes with different tilting effects in LSCO. The insets show the tilting effects in accordance with the Glazer's notations.~\cite{Glazer2011}}
\label{Fig:1}
\end{figure}

\begin{figure*}
\includegraphics{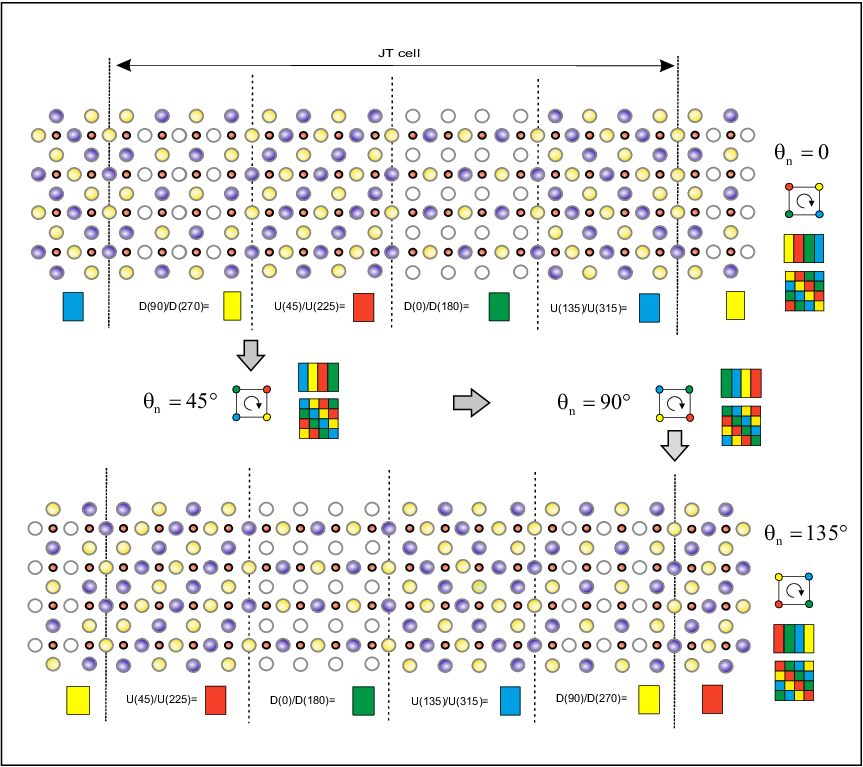}
\caption{\label{fig:2} Structural motive for the four-fold degenerate state  $\Psi_{\theta_n}$ of the novel JT cell. The initial phase $\theta_n$, $\Psi_{\theta_n}$ - color code  and coloring of the graph with $\chi=2$ are also shown.}
\label{Fig:2}
\end{figure*}

We focus on the tilting effects as a required attribute of observed stripes~\cite{Bianconi1996}. The idea of the color approach~\cite{Gavrichkov2019} is the identification of stripes due to the tilting effects (tilting of the CuO$_6$ octahedron as a whole). The fact is that different stripes differ not only in the hole concentration and spin, but in the type of nuclear configuration (see Fig.\ref{Fig:1})

We have classified the observed stripe configurations into a ninth-order symmetric Abelian group $G(\alpha)$ consisting of two types of stripes $U(\theta_U)$ and $D(\theta_D)$ rotated at the right angle relative to each other.~\cite{Gavrichkov2019} The sought stripe group $G(\alpha)$  represents all possible reaction products between different initial nuclear configurations in the form of group multiplication, where an ideal non-tilted prototype of the perovskite CuO$_6$  octahedron is used as a group unit. In all the "structural reactions" of nuclear configurations of the stripes,  the tilting angle
$\alpha$ of the perovskite octahedron with respect to the rock salt LaO layers is retained, rather than their initial symmetry, as is the case, for example, of chemical reactions, in accordance with the Woodworth-Hoffman rules.~\cite{Woodward1971}  The number of nuclear configurations of the observed stripes in LSCO, is reduced to one of the two $U(\theta_U)$  (where $\theta_U=45^\circ,135^\circ,225^\circ,315^\circ$ ) or $D(\theta_D)$  (where $\theta_D=0,90^\circ,180^\circ,270^\circ$) stripes, and their spatial distribution in the CuO$_2$ layer is represented by possible plane graphs with the chromatic number $\chi  \le 4$ in the four-color theorem.  Using four colors: $U(45^\circ),U(225^\circ)$  - red ($R$),  $U(135^\circ),U(315^\circ)$ - blue ($B$) and also $D(0^\circ),D(180^\circ)$ - green ($G$); $D(90^\circ),D(270^\circ)$  - yellow ($Y$), we can always color an arbitrary plane map, and the four colors $R$, $G$, $B$, $Y$ are just four subgroups in $G(\alpha)$. The group can also be represented as a direct product of the subgroups of any different colors, e.g. $G(\alpha)=R\times B=G\times Y$, etc. The $D(\theta_D)$ stripes occur as overlapping different $U(\theta_U)$  and    $U(\theta_U\pm90^\circ)$ stripes in an $R/G/B$ sequence, for example, upon $U(\theta_U)$  stripes disordering with doping, and vice versa, $U(\theta_U)$  stripes can occur as overlapping $D(\theta_D)$   and $D(\theta_D\pm90^\circ)$  stripes in the $G/R/Y$ sequence.

These two stripe sequences can be combined into one, for example into $Y/R/G/B$, from which, by rotating all the tilted CuO$_6$ octahedra around the $c$ axis by the angle $\delta \theta  = n \cdot \left( {45^\circ } \right)$, we can obtain other three $B/Y/R/G$, $G/B/Y/R$ and $R/G/B/Y$.
The rotation also leads to a shift over the $G(\alpha)$ group and  transverse shift of the linear stripe and diagonal shift of the checkerboard structures in Fig.\ref{Fig:2}. This means that the ground state of the CuO$_2$  layer in Fig.\ref{Fig:2} is fourfold degenerate with respect to the rotation. Indeed, we can also form a novel JT cell (see Fig.\ref{Fig:2}), and during its translation the number of hole carriers and spins in the JT cell is retained. The state  $\Psi _{\theta _n }$ of the JT cell is fourfold degenerate by the initial phase $\theta _n  = 0,45^\circ ,90^\circ ,135^\circ$, which splits upon tunneling the CuO$_6$ octahedra over the states of the $D\left( {\theta _D } \right)$ and $U\left( {\theta _U } \right)$ stripes. With the spontaneous $\theta$-symmetry breaking, a static picture which is formed with the help of the JT cell is shown in Fig.\ref{Fig:2}.

\begin{figure*}
\includegraphics{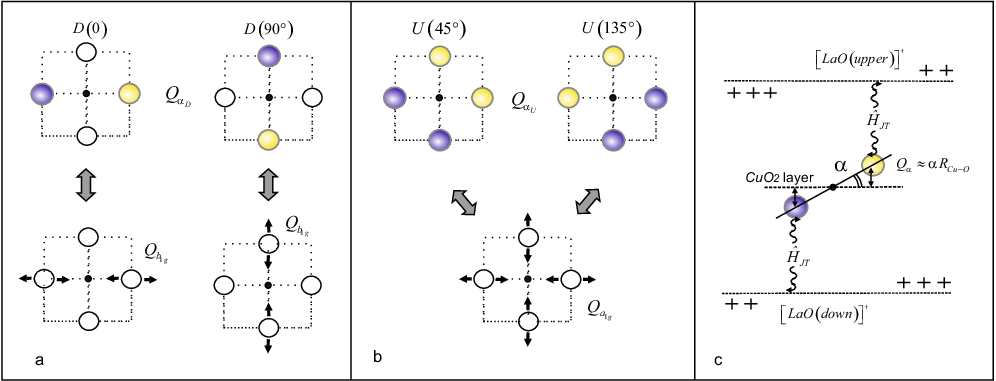}
\caption{\label{fig:3}  Graphic scheme: (a) $b_{1g}$  tilting modes active in the non-local CuO$_6$  octahedron JT effect, as well as (b)  $a_{1g}$ relaxation modes in the $D$ and $U$ stripes. The arrows in (a) and (b) show the relationship between the tilting and conventional local modes. (c) Non-local JT effect in the CuO$_2$ layer surrounded by the symmetrical LaO rock salt layers.}
\label{Fig:3}
\end{figure*}

Let us consider the JT cell to understand the nature of the inhomogeneous distribution of the hole and spin density in the CuO$_2$  layer. The cell has the Jahn-Teller (JT) nature, as shown in Fig.\ref{Fig:3}, where the (pseudo)$JT$ effect is non-local, and associated with the presence of positively charged LaO  rock salt layers. The tilting angle $\alpha_U$ at the orientation angle $\theta_U$   corresponds to the relaxation  $a_{1g}$ modes, and the tilting angle $\alpha_D$  at the orientation angle $\theta_D$  corresponds to the JT  active $b_{1g}$ modes.

The absence of superconductivity in the AFM CuO layers on graphene ~\cite{Kano2017, Kvashnin2019} indicates the key role of rock salt layers in $2D$ doped cuprates.~\cite{Bianconi2001}  The direct similarity of the structures of the undoped LCO and non-JT La$_2$NiO$_4$ materials, ~\cite{Rodriguez1991} where Ni$^{2+}$ ions are in the high spin state $S=1$, which also indicates the missing JT effect in the $N_0(d^9)$ sector of the configuration space of the CuO$_2$ layer. However, in the $N_-(d^8)$ sector of the doped LSCO, the hole carriers  are in the Zhang-Rice singlet state $^1A_{1g}$,~\cite{Zhang1988}, this being a JT state.~\cite{Bersuker1992} Moreover, the JT pseudo-effect depends on the doping concentration in a threshold way. Indeed, the hole carriers in the  $D(\theta_D)$ stripes can induce a non-local JT pseudo-effect in the $i$-th CuO$_6$ octahedron with a Hamiltonian (see Fig.\ref{Fig:3})
\begin{widetext}
\begin{equation}
\hat H^{(i)}_{JT} \left( {Q_\alpha  ,Q_\theta  } \right) = \varepsilon _{{}^1B_{1g} } \sum\limits_\sigma  {b_{i\sigma }^ +  b_{i\sigma }  + } \varepsilon _{{}^1A_{1g} } \sum\limits_\sigma  {a_{i\sigma }^ +  a_{i\sigma }  + } V_{b_{1g} } Q_\alpha \sum\limits_\sigma  {\left( {a_{i\sigma }^ +  b_{i\sigma }  + b_{i\sigma }^ +  a_{i\sigma } } \right)}  + U\left( {Q_\alpha ,Q_\theta  } \right)
\label{eq:1}
\end{equation}
\end{widetext}

in the hole sector $N_-(d^8)$ of the configuration space of the  CuO$_2$ layer.
Here, $a_{i\sigma }^ +   = \eta \left( \sigma  \right)X_i^{{}^1B_{1g} ,\bar \sigma _b }$ and $b_{i\sigma }^ +   = \eta \left( \sigma  \right)X_i^{{}^1A_{1g} ,\bar \sigma _b }$
are two hole parts of the quasiparticle operator
\begin{eqnarray}
&&d_{i\lambda \sigma }^ +   =  =X_i^{\sigma _b ,A_{1g} }  + \sum\limits_{h = {}^1A_g ,{}^1B_g } {\left\langle h \right|} d_{i\lambda \sigma }^ +  \left| {\sigma _b } \right\rangle X_i^{h,\bar \sigma _b }  =
\\ \nonumber
&&=X_i^{\sigma _b ,A_{1g} }  + \left\{ {\begin{array}{*{20}c}
   {a_{i\sigma }^ +  ,\lambda  = 3z^2  - r^2 }  \\
   {b_{i\sigma }^ +  ,\lambda  = x^2  - y^2 }  \\
\end{array}} \right.
\label{eq:2}
\end{eqnarray}
generating two-hole $^1A_{1g}$  and  $^1B_{1g}$ states in the  $N_-(d^8)$ sector. If $\left| h \right\rangle  = \left| {{}^1A_{1g} } \right\rangle$, then the operator $d_{i\lambda \sigma }^ +$
 takes the well-known form for the Hubbard model.~\cite{Hubbard1964} It is assumed that the $U(Q_\alpha,Q_\theta)$ potential  is created by the  LaO rock salt layers, and the equilibrium ion positions in the CuO$_2$  layer can be obtained from the equation  $\frac{{\partial \hat H_{JT} \left({Q_\alpha,Q_\theta} \right)}}{{\partial Q_\alpha  \partial Q_\theta  }} = 0$. Here, we just use the set $\theta_D$   and $\alpha_D\approx14^\circ\div18^\circ$, which are observed  at $T=100K$  for the $D$  stripes.~\cite{Bianconi1996}
The concentration dependence of the magnitude  $\alpha_D(x)$ in the $JT$ effect  can be obtained using the $u/v$ transformation in the Hamiltonian (\ref{eq:1}) with the coefficients
\begin{eqnarray}
&&u^2 \left( {Q_\alpha  } \right) = \frac{1}{2}\left( {1 - \frac{{\varepsilon _{{}^1B_{1g} } - \varepsilon _{{}^1A_{1g} } }}{{D\left( {Q_\alpha  } \right)}}} \right),\\ \nonumber
&&v^2 \left( {Q_\alpha  } \right) = \frac{1}{2}\left( {1 + \frac{{\varepsilon _{{}^1B_{1g} } - \varepsilon _{{}^1A_{1g} } }}{{D\left( {Q_\alpha  } \right)}}} \right),
\label{eq:3}
\end{eqnarray}
where $D\left( {Q_\alpha  } \right) = \sqrt {\left(  \varepsilon _{{}^1B_{1g} }-{\varepsilon _{{}^1A_{1g} } } \right)^2  + 4\left( { V_{b_{1g}}Q_\alpha  } \right)^2 }$ and $a_{i\sigma }^ +   = v\left( {Q_\alpha  } \right)\tilde a_{i\sigma }^ +   + u\left( {Q_\alpha  } \right)\tilde b_{i\sigma }^ +$, $b_{i\sigma }^ +   =v\left( {Q_\alpha  } \right)\tilde b_{i\sigma }^ - - u\left( {Q_\alpha} \right)\tilde a_{i\sigma }^ +  $. Then, Hamiltonian (\ref{eq:1}) may be written as

\begin{equation}
\hat H^{(i)}_{JT} \left( {Q_\alpha  ,Q_{\theta _D } } \right) = \varepsilon _ -  \left( {Q_\alpha  } \right)\sum\limits_\sigma  {\tilde b_{i\sigma }^ +  \tilde b_{i\sigma }^{} }  + \varepsilon _{_ +  } \left( {Q_\alpha  } \right)\sum\limits_\sigma  {\tilde a_{i\sigma }^ +  \tilde a_{i\sigma }^{} },
\label{eq:4}
\end{equation}
where $\varepsilon _ \pm  \left( {Q_\alpha  } \right) = \frac{1}{2}\left[ {\varepsilon _{{}^1A_{1g} }  + \varepsilon _{{}^1B_{1g} }  \pm D\left( {Q_\alpha } \right)} \right]$
are the energies of the hole quasiparticles in the two-hole singlets  $\left| {{}^1\tilde A_{1g} } \right\rangle$ and $\left| {{}^1\tilde B_{1g} } \right\rangle$ taking into account the JT interaction $V_{b_{1g} }$.
The equilibrium tilting angle $\alpha_D$   at  $U\left( {Q_\alpha  ,Q_{\theta _D } } \right) \approx {{\left( {K  \alpha _{}^2 } \right)} \mathord{\left/
 {\vphantom {{\left( {K  \alpha _{}^2 } \right)} 2}} \right.
 \kern-\nulldelimiterspace} 2}$ takes the form
$\alpha_D(x)  =  \pm \sqrt {\left( {\frac{{R V_{b_{1g} } }}{{K }}} \right)^2  - \left( {\frac{\Delta }{{V_{b_{1g} } }}} \right)^2 }$
where $ R  = \sum\limits_\sigma  {\left\langle {\tilde b_\sigma ^ +  \tilde b_\sigma   - \tilde a_\sigma ^ +  \tilde a_\sigma  } \right\rangle }$.  Thus, in the doped LSCO cuprates there is a hole doping level  $x_c  = \left\langle {X_i^{{}^1\tilde A_{1g} {}^1\tilde A_{1g} }  - X_i^{{}^1\tilde B_{1g} {}^1\tilde B_{1g} } } \right\rangle  = \frac{{K  \Delta }}{{V_{b_{1g} }^2 }}$
, and above which, the JT pseudo effect with $\alpha_D  \ne 0$ can be observed.  At  $\Delta  = \varepsilon _{{}^1B_{1g} }  - \varepsilon _{{}^1A_{1g} }  = 0$ ~\cite{Gavrichkov1990} we obtain the quadratic hole doping dependence for the JT contribution:
\begin{equation}
\hat H^{(i)}_{JT} \left( {Q_{\alpha _D } ,Q_{\theta _D },x } \right) =  - E^{(0)}_{JT} R\sum\limits_\sigma  {\left( {\tilde b_{i\sigma }^ +  \tilde b_{i\sigma }^{}  - \tilde a_{i\sigma }^ +  \tilde a_{i\sigma }^{} } \right)},
\label{eq:5}
\end{equation}
where $E^{(0)}_{JT}  = {{V_{b_{1g} }^2 } \mathord{\left/ {\vphantom {{V_{b_{1g} }^2 } {2K  }}} \right. \kern-\nulldelimiterspace} {2K  }}$. Any homogeneous hole density at $x < x_c$ will be unstable to the creation of local $D$ areas with a higher hole density $x_D > x_c$ and $\left\langle {\mathord{\buildrel{\lower3pt\hbox{$\scriptscriptstyle\frown$}}
\over H} _{JT} } \right\rangle  = \left\langle {\sum\limits_{i = i_D } {\mathord{\buildrel{\lower3pt\hbox{$\scriptscriptstyle\frown$}}
\over H} _{JT}^{(i)} } } \right\rangle  \ne 0$. The boundaries contribution to the lattice energy is absent, because of the boundaries between the stripes are in fact atomically sharp.
At the small hole doping, the total JT contribution $\left\langle {\mathord{\buildrel{\lower3pt\hbox{$\scriptscriptstyle\frown$}}
\over H} _{JT} } \right\rangle \sim x$ due to the increasing width of the $D$ stripes with the constant hole concentration $x_D$. This is a $x_D$ - constant scenario, where $x_D\geq x_c$.

\section{\label{sec:III} Electron-hole pairs in the superexchange interaction \\}
\begin{figure}
\includegraphics{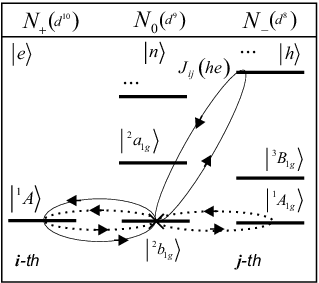}
\caption{\label{fig:4} Graphic representation of the electron-hole pairs for the interacting $i$ -th and $j$ - th Cu$^{2+}$ ions. Each pair corresponds to a double exchange loop (solid or dotted lines)}
\label{Fig:4}
\end{figure}

3.1 Let us consider the superexchange interaction $J_{ij}$ in the static stripe structure in Fig.\ref{Fig:2}. In the case of spontaneous $\theta$-symmetry breaking (the JT cell state $\Psi_{\theta_n}$  is non-degenerate), the doped LSCO cuprate has a static nanostructure, where the exchange interaction is missing in the $D$  stripes (where $S_i=0$), and in the  $U$ stripes the superexchange interaction is equal to the one in undoped LCO materials:

\begin{equation}
\hat H_S  =  - \sum\limits_{ij} {J_{ij}^{tot} \hat S_i \hat S_j }
\label{eq:6}
\end{equation}
 Here, the interaction  $J_{ij}^{tot}  = \sum\limits_{h,e} {J_{ij} \left( {h,e} \right)}$ can be calculated in the adiabatic approximation. Superexchange interaction (\ref{eq:6}) arises as a result of the superposition of contributions from all possible virtual electron-hole pairs on the interacting $i$-th and $j$-th  ions   (see Fig.\ref{Fig:4}).
The sign of the  $J_{ij} \left( {h,e} \right)$ contribution to the $J_{ij}^{tot}$ interaction is determined by a simple rule. If the spins of an electron and a hole in a virtual pair are equal   , $\hat S_h  = \hat S_e$, this is the AFM contribution. If $\hat S_h  = \hat S_e  \pm 1$ it is the FM contribution. Other terms are missing, and all the electron-hole wave functions $\left| h \right\rangle$ and $\left| e\right\rangle$  in the hopping integral  $t_{ij}^{hn,en}  = \sum\limits_{\lambda \lambda '} {t_{ij}^{} \left( {\lambda \lambda '} \right)\sum\limits_\sigma  {\gamma _{\lambda \sigma }^* \left( {hn} \right)\gamma _{\lambda '\sigma } \left( {en} \right)} }$  ~\cite{Gavrichkov2020, Gavrichkov2017, Mikhaylovskiy2020} in the superexchange

\begin{eqnarray}
&&J_{ij}^{tot}  = \sum\limits_{h,e} {J_{ij} \left( {hn,en} \right)},
J_{ij} \left( {hn,en} \right) = \frac{{2\left( {t_{ij}^{hn,en} } \right)^2 }}{{\Delta \left( {hn,en} \right)}}, \\ \nonumber
&&\Delta \left( {hn,en} \right) = \varepsilon _h  + \varepsilon _e  - 2\varepsilon _n
\label{eq:7}
\end{eqnarray}
are calculated in the adiabatic approximation, where

\begin{equation}
\begin{array}{l}
 \gamma _{i\lambda \sigma } \left( {hn} \right) = \left\langle h \right|d_{i\lambda \sigma }^ +  \left| n \right\rangle  = \left\langle {\left( {N_ -  ,S_h } \right)} \right|d_{i\lambda \sigma }^ +  \left| {\left( {N_0 ,S_n } \right)} \right\rangle  \\
 \gamma _{i\lambda \sigma } \left( {en} \right) = \left\langle e \right|d_{i\lambda \sigma }^{} \left| n \right\rangle  = \left\langle {\left( {N_ +  ,S_e } \right)} \right|d_{i\lambda \sigma }^{} \left| {\left( {N_0 ,S_n } \right)} \right\rangle  \\
 \end{array}
\label{eq:77}
\end{equation}

The pairs of indices $hn$  and $en$  run over all possible quasiparticle excitations  $\left( {e,n} \right)$ and $\left( {h,n} \right)$  between many-electron states  $\left| n \right\rangle$ and  $\left| {e\left( h \right)} \right\rangle$ with the energies  $\varepsilon _n$ and $\varepsilon _{e(h)}$  in the sectors $N_0$ and $N_ \pm$ of the configuration space in Fig.\ref{Fig:4}.

These quasiparticle excitations are described by non-diagonal elements $t_{ij}^{nh,ne}$.  In the Hubbard's model there is only one such element corresponding to the excitations between lower and upper Hubbard bands. In cuprates, the main contribution from the electron-hole pair with $\left| n \right\rangle  = {}^2b_{1g}$, $\left| h \right\rangle  = {}^1A_{1g}$ and $\left| e \right\rangle  = {}^1A$  has the AFM character $J_{ij}^{tot}  \approx J_{ij}^{\left( U \right)} \left( {{}^1A{}^2b_{1g} ,{}^1A_{1g} {}^2b_{1g} } \right)$ (see Fig.\ref{Fig:4}), ~\cite{Gavrichkov2016} where
\begin{equation}
J_{ij}^{\left( U \right)} \left( {{}^1A{}^2b_{1g} ,{}^1A_{1g} {}^2b_{1g} } \right) = \frac{{2t_{ij}^2 }}{{\Delta \left( {{}^1A{}^2b_{1g} ,{}^1A_{1g} {}^2b_{1g} } \right)}} \approx 0.15eV
\label{eq:8}
\end{equation}

where $\Delta \left( {{}^1A{}^2b_{1g} ,{}^1A_{1g} {}^2b_{1g} } \right) = \varepsilon _{{}^1A_{1g} }  - 2\left( {\varepsilon _{{}^2b_{1g} }  - V_{a_{1g} } \alpha _U } \right) = U_H$ and $\varepsilon _{{}^1A_{1g} }  = 2\left( {\varepsilon _{{}^2b_{1g} }  - V_{a_{1g} } \alpha _U } \right) + U_H$. However, the transverse percolation of the exchange interaction $J_{i_R j_B }^{tot}$, where $i_R$ and   $j_B$ in the octahedra CuO$_6$ are in different red($R$) and blue($B$) stripes in the static regular $Y/R/G/B...$ stripe structure, is unlikely.

\begin{figure*}
\includegraphics{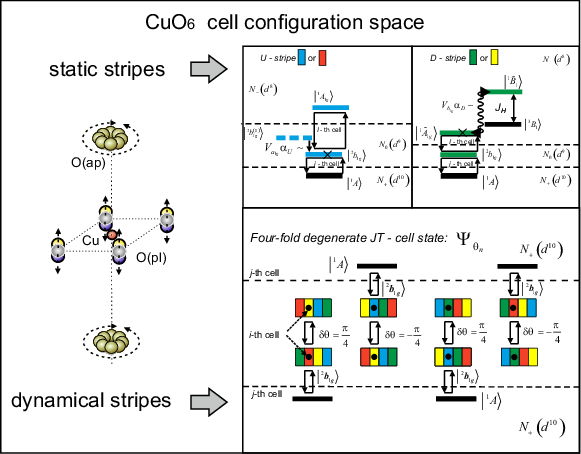}
\caption{Left graphical representation shows the anisotropic non-local effects in the CuO$_6$ octahedron located in the JT cell with the fourfold degeneracy. The right diagram presents the electron-hole pairs contributing to the exchange interaction when both the $i$ -th and $j$ - th magnetic ions are in the JT cell. Here, the $U$ and $D$  stripe widths are the same to demonstrate the appearance of an ideal CuO$_2$   layer with mobile hole carriers only with the equal stripe widths. }
\label{Fig:5}
\end{figure*}

3.2 If there is no spontaneous $\theta$-symmetry breaking and the JT cell state is four-fold degenerate, then the exchange interaction in the JT cell presented above is incomplete,
due to the the novel configuration space.
Here, the $i$-th CuO$_6$ octahedron
(the black point in Fig.\ref{Fig:5}) can simultaneously be in the  $U$ and  $D$ stripe states
of any color of the four possible $R,B$ and $G,Y$.
In addition to the initial quasiparticles in the static stripe structure, specific hole quasiparticles appear in the CuO$_2$ layer, but they are accompanied by changes $\delta \alpha=\alpha_D-\alpha_U$ and $\delta \theta=\theta_D-\theta_U$
shown in Fig.\ref{Fig:5})

To obtain contributions into $J_{ij} \left( {hn,en} \right)$ from the electron-hole pairs with a change in the angles $\delta \alpha$ and $\delta \theta  =  \pm 45^\circ$ in the JT cell,  we need the $i$-th CuO$_6$ octahedron  $\left| {h_{\theta _D \alpha_D } } \right\rangle$ and $\left| {n_{\theta _U \alpha_U } } \right\rangle$  states, where the indices $U$  and $D$  denote the stripe affiliation for the $i$-th octahedron
\begin{equation}
\left| {h_{\theta _D \alpha _D } } \right\rangle  = \left| {\tilde h} \right\rangle \left| {\chi _{\theta _D }^{} \chi _{\alpha_D }^{} } \right\rangle; \left| {n_{\theta _U \alpha_U } } \right\rangle  = \left| n \right\rangle \left| {\chi _{\theta _U }^{} \chi _{\alpha_U }^{} } \right\rangle,
\label{eq:9}
\end{equation}
where $\left| {\tilde h} \right\rangle  = \left| {{}^1\tilde A_{1g} } \right\rangle ,\left| {{}^1\tilde B_{1g} } \right\rangle$ and $\left| n \right\rangle  = \left| {{}^2b_{1g} } \right\rangle$. Each of the $R,G,B,Y$ colors in Fig.\ref{Fig:2} corresponds to a harmonic oscillator wave function   $\left| {\chi _\theta ^{} } \right\rangle$ with a displaced $2D$ oscillator. These states differ by  $\left| {\chi _{\alpha_D } \left( {Q_\alpha  + \alpha_D } \right)} \right\rangle$, $\left| {\chi _{\theta _D } \left( {Q_{\theta _D } } \right)} \right\rangle$ and    $\left| {\chi _{\alpha _U } \left( {Q_\alpha + \alpha_U } \right)} \right\rangle$, $\left| {\chi _{\theta _U } \left( {Q_{\theta _U } } \right)} \right\rangle$, and the matrix elements $\gamma _{\lambda \sigma } \left( r \right)$ in Eq.(\ref{eq:77}) for the $i$-th CuO$_6$ octahedron cell will be calculated with the functions $\left| {h_{\theta _D \alpha _D } } \right\rangle$ and   $\left| {n_{\theta _U \alpha _U } } \right\rangle$ located at the four equivalent nuclear configurations of the JT cell  corresponding to the phase $\theta _n$  (see black dot in Fig.\ref{Fig:5}). The matrix elements   $\gamma _{\lambda \sigma }^{} \left( {hn} \right)$ contain vibronic reduction factors or partial dynamic quenching:~\cite{Ham1965}
\begin{eqnarray}
&&J_{ij} \left( {{}^1\tilde A_{1g} ,{}^1A } \right) = \frac{{2\left( {t_{ij}^{{}^1\tilde A_{1g} {}^2b_{1g} ,{}^1A {}^2b_{1g} } } \right)^2 }}{{\Delta \left( {{}^1\tilde A_{1g} {}^2b_{1g} ,{}^1A {}^2b_{1g} } \right)}} \approx \\ \nonumber
&&\approx\frac{{2v^2 \left( {\alpha _D } \right) \cdot t_{ij}^2 \left( {x^2 ,x^2 } \right)}}{{\Delta \left( {{}^1\tilde A_{1g} {}^2b_{1g} ,{}^1A {}^2b_{1g} } \right)}} \cdot \exp \left\{ { - \nu\frac{{\delta \theta _{}^2  + \delta \alpha_{}^2 }}{2}} \right\},
\label{eq:10}
\end{eqnarray}
where $\nu=K/\hbar\omega_D$ ($\omega_D$ - Debye frequency) and 
\begin{widetext}
\begin{equation}
t_{ij}^{{}^1\tilde A_{1g} {}^2b_{1g} ,{}^1A {}^2b_{1g} }  \approx t_{ij}^{} \left( {x^2 ,x^2 } \right)v\left( {\alpha _D } \right)\sum\limits_\sigma  {\gamma _{x^2 \sigma }^* \left( {{}^1A_{1g} ,{}^2b_{1g} } \right)\gamma _{x^2 \sigma } \left( {{}^1A ,{}^2b_{1g} } \right)}  \cdot \left\langle {\chi _{i\theta _D }^{} } \right|\left. {\chi _{i\theta _U }^{} } \right\rangle \left\langle {\chi _{i\varphi _D }^{} } \right|\left. {\chi _{i\varphi _U }^{} } \right\rangle ,
\label{eq:11}
\end{equation}
\end{widetext}
where $\Delta \left( {{}^1\tilde A_{1g} {}^2b_{1g} ,{}^1A {}^2b_{1g} } \right) = 2\left[ {\varepsilon _{_ -  } \left( {\alpha _D } \right) - \varepsilon _{{}^2b_{1g} } } \right] + U_H$  ,    $t_{ij}^{} \left( {x^2 ,x^2 } \right) \gg t_{ij}^{} \left( {z^2 ,z^2 } \right)$, and
\begin{equation}
\begin{array}{l}
 \left\langle {\chi _{i\theta _D }^{} } \right|\left. {\chi _{i\theta _U }^{} } \right\rangle  \cdot \left\langle {\chi _{i\alpha _D }^{} } \right|\left. {\chi _{i\alpha _U }^{} } \right\rangle  =  \\
   \left\langle {\chi _{i\alpha _D } \left( {Q_\alpha   + \alpha _D } \right)\chi _{i\theta _D } \left( {\theta _D } \right)} \right|\left. {\chi _{i\alpha _U } \left( {Q_\alpha  + \alpha _U } \right)\chi _{i\theta _U } \left( {\theta _U } \right)} \right\rangle \\
  \approx \exp \left\{ { - \nu{{\left( {\delta \theta _{}^2  + \delta \alpha _{}^2 } \right)} \mathord{\left/
 {\vphantom {{\left( {\delta \theta _{}^2  + \delta \alpha _{}^2 } \right)} 4}} \right.
 \kern-\nulldelimiterspace} 4}} \right\} \\
 \end{array}
\label{eq:13}
\end{equation}
 with $\delta \alpha  = \alpha_D  - \alpha_U  \approx 9^\circ \div 13^\circ$. With the absence of spontaneous $\theta$-symmetry breaking, the homogeneous superexchange interaction in the hole-doped   LSCO cuprates is equal to
\begin{equation}
J_{ij}^{tot}  \approx J_{ij} \left( {{}^1\tilde A_{1g} {}^2b_{1g} ,{}^1A_1 {}^2b_{1g} } \right) \cdot v^2\left( {\alpha_D } \right) \cdot \exp \left\{ { -\nu\frac{{\delta \theta _{}^2  + \delta \alpha_{}^2 }}{2}} \right\}.
\label{eq:14}
\end{equation}
The result (\ref{eq:14})  is independent of the choice of the $i$ and $j$ pair of the interacting   CuO$_6$ octahedra and the exchange $J_{ij}^{tot}$  is homogeneous in the ideal CuO$_2$ layer with nonlocal effects. Indeed, the neutron experiments in LCO  cuprates demonstrate the thermal anisotropic motion ~\cite{Hafliger2014} similar to the oxygen ion motion shown in Fig.\ref{Fig:5}. Similar reduction factors in $n$ doped  LNCO cuprates are impossible, since there is no JT effect in the $N_ +  \left( {d^{10} } \right)$ electron sector.

\section{\label{sec:VI}Discission and conclusions  \\}
The JT pseudo effect in the $N_ -  \left( {d^8 } \right)$ hole sector with the hole concentration $x$ in the range $0 < x < x_c$ is accompanied by the charge and spin inhomogeneities due to the nonzero tilting angle  $\alpha_D  \approx 14 \div 18^\circ$ ~\cite{Bianconi1996} of the CuO$_6$ octahedra  at the orientation angle $\theta _D=0^\circ, 90^\circ, 180^\circ, 270^\circ$ in the $D\left( {\theta _D } \right)$ stripes (see Fig.\ref{Fig:1}). The $U\left( {\theta _U } \right)$ stripes with the CuO$_6$ octahedra  tilted at an angle $\alpha _U  \approx 5^\circ$~\cite{Pickett1989} and oriented at an angle  $\theta_U=45^\circ,135^\circ,225^\circ,315^\circ$ form dielectric regions.
The observed regular line and checkerboard stripe structures~\cite{Seibold2007, Okamoto2012} with the chromatic number $\chi  = 2$ generate a novel element of symmetry in the CuO$_2$ layer: simultaneously rotating all the tilted $CuO_6$  octahedra by the angle $\delta \theta_n  = n \cdot \left( {45^\circ } \right)$ (where $n=1\div4$) around the $c$ axis. We can choose a novel JT cell, with the initial structure and number of hole carriers and spins in the JT cell being retained during its translation. The state $\Psi _{\theta _n }$ of the JT cell is fourfold degenerate by the initial phase $\theta _n  = 0,45^\circ ,90^\circ ,135^\circ$ (see Fig.\ref{Fig:2}).
In particular, any CuO$_6$ octahedron can be located simultaneously, both in the $U\left( {\theta _U } \right)$ and $D\left( {\theta _D } \right)$ stripes without any well-specified orientation $\theta _D \left( {\theta _U } \right)$ and tilting $\alpha _D \left( {\alpha _U } \right)$ angles. The exchange interaction $J_{ij}^{tot}$ will be different, depending on whether the spontaneous  phase $\theta$-symmetry breaking occurs or not:\\

(i) The spontaneous $\theta$-symmetry breaking leads to a static spatial distribution of the $U\left( {\theta _U } \right)$ and $D\left( {\theta _D } \right)$ stripes in the doped LSCO. The static distribution origins from the different stripe width and shape, when there is no $ \theta_n$ phase degeneracy. The signature of static structure do not depend on experimental timescale ranging from $10^{-6}$s to $10^{-15}$s, but is sensitive to temperature and hole concentration~\cite{Lanzara1999}. In the static structure of the CuO$_2$ layer, the superexchange interaction has the Anderson's form   $J_{ij}^{\left( U \right)}  = {{2t^2 } \mathord{\left/
 {\vphantom {{2t^2 } {U_H }}} \right.
 \kern-\nulldelimiterspace} {U_H }}$ for the interacting ions  Cu$^{2 + }$ with spin ${1 \mathord{\left/
 {\vphantom {1 2}} \right.
 \kern-\nulldelimiterspace} 2}$ , but is limited in space  by the $U\left( {\theta _U } \right)$ stripes with the zero hole concentration.

(ii) Without spontaneous $\theta$-symmetry breaking a novel JT cell can be constructed in the periodic stripe nanostructure with the fourfold degenerate $\Psi _{\theta _n }$ state. The ideal $CuO_2$   structure with nonlocal anisotropic  effects and the homogeneous superexchange are restored. The stripe boundaries, spin and charge inhomogeneities disappear,  and the superexchange is suppressed from its magnitude $J_{ij}^{\left( U \right)}$ (in the undoped $LCO$ cuprate) by the exponential factor $J_{ij}^{tot}  \approx J_{ij}^{\left( U \right)}  \cdot v^2\left( {\alpha _D } \right) \cdot \exp \left\{ { - \nu{{\left( {\delta \alpha _D^2  + \delta \theta _D^2 } \right)} \mathord{\left/
 {\vphantom {{\left( {\delta \alpha _D^2  + \delta \theta _D^2 } \right)} 2}} \right.
 \kern-\nulldelimiterspace} 2}} \right\}$ due to dynamic quenching for the CuO$_6$ octahedra. 
We can evaluate the $E_{JT}$ magnitude as $E_{JT}  \approx \hbar \omega _D$, where $E_{JT}  = K\left( {\delta \alpha ^2  + \delta \theta ^2 } \right)$, because the stripe fluctuations are observed at the temperature $\sim 100K$. Thus  $J_{ij}^{tot}  \approx  0.61  \cdot J_{ij}^U  \cdot v^2 \left( {\alpha _D } \right)$, where
$v^2 \left( {\alpha _D }\right)\approx 1$ and $J_{ij}^U \approx 0.15eV$. The strip signatures noticeably weaken with decreasing temperature, but are still detected as density waves of hole pairs~\cite{Li2019, Du2020, Tsvelik2019, Chen2022}.

There is no direct analogy with a free rotator, since there is no continuous series $\theta_D$ of the tilted CuO$_6$ octahedra for which the potential energy is minimal. Only tunneling and hopping effects are possible at the low and high temperatures, respectively. This conclusion does not contradict the results, where the low-temperature optical pump, soft x-ray probe measurements ~\cite{Mitrano2019} detected the unexpected gapless nature of the charge ordering. Indeed, the characteristic tunneling JT splitting $\le 10sm^{ - 1} \left( {1meV} \right)$ ~\cite{Bersuker2010} has the energy scale of the transverse fluctuations observed in $LBCO$~\cite{Mitrano2019}, and the magneto-optical measurements identify the effects of time reversal symmetry breaking~\cite{Xia2006, Xia2008, Li2011, He2011}.


Note that a period of the charge stripes sequence $U/D/U/...$ is two times less than the period of the lattice stripe structure $Y/R/G/B...$, since the hole concentration $x_D$ does not depend on the orientation angle $\theta _D$. In (nonpseudo) Jahn-Teller materials, where $x_c=0$ such as 3D AMeO$_3$ perovskites  with regular $MeO_6$ octahedra, the hole segregation into strip structures is impossible. The hole concentration $x_D$ in the $D$ stripes does not depend on the total hole concentration $x$ in the CuO$_2$ layer, but the total $D$ stripe area rises with increasing hole doping. This $x_D$ constant scenario has clear experimental features such as Fermi level pinning, where, a shift of chemical potential  is suppressed in the underdoped region of $p$ type LSCO cuprates ~\cite{Harima2001}. In the $n$ type LNCO cuprates, the JT effect in $N_ +  \left( {d^{10} } \right)$ configuration sector is missing. In contrast to LSCO the superexchange interaction $J_{ij}^{}$ should not be subject to dynamic quenching, and the monotonous increase of the chemical potential is consistent with the absence of stripe fluctuations ~\cite{Harima2001}. The exchange interaction in Eq.(\ref{eq:14}) decreases exponentially with increasing Debye temperature ${{\hbar \omega _D } \mathord{\left/
 {\vphantom {{\hbar \omega _D } {k_B }}} \right.
 \kern-\nulldelimiterspace} {k_B }}$. Indeed, similar trend for the critical temperature $T_C$ in the  La, Y, Tl and Hg based cuprates with Debye temperatures from $200K$ to $600K$ was found in the works ~\cite{Ledbetter1991, Ledbetter1994}. Note also our arguments in favor of the JT nature of the charge inhomogeneity is an alternative to the fact that the competition between the kinetic energy and Coulomb repulsion could cause holes to segregate into strip structures~\cite{Imada2021}.

We are grateful to the colleagues who made useful comments after the draft paper in arxiv. The work was supported by the Russian Science Foundation, research grant RSF No.22-22-00298.

\bibliography{paper_gavrichkov}
\end{document}